\documentclass[final,1p,times]{elsarticle}
\usepackage{amssymb}
 \usepackage{amsmath}
 \usepackage{url}
\usepackage{color}
\usepackage{lineno}

\journal{ }
\date{\today}
\begin{document}  
\title{Transverse coupled-bunch instabilities driven by high-order modes near the coupling resonance }

\author{You Sun }
\author{Weiwei Li\corref{1}}
\author{Tianlong He }
\author{Penghui Yang }
\author{Xiaoyu Liu }
\author{Zhenghe Bai}

\cortext[1]{Corresponding author. Email: liwe@ustc.edu.cn}

\address{National Synchrotron Radiation Laboratory, University of Science and Technology of China, Hefei, Anhui, 230029, China}

\begin{abstract}
Betatron coupling near the difference resonance has been explored and adopted in several fourth-generation storage rings, yet its influence on high-order-mode-driven transverse coupled-bunch instabilities has not been systematically established. Using the Hefei Advanced Light Facility (HALF) as an example, we investigate this effect through theoretical analysis and macroparticle tracking simulations. The theoretical predictions agree well with macroparticle tracking simulations in both the weak- and strong-instability regimes, validating the description over the parameter ranges considered. For the vertical HOMs studied with HALF parameters, suitable betatron coupling effectively suppresses the instability, offering potential benefits for transverse stability control.
\end{abstract}


\maketitle
\section{Introduction}
\label{sec:introduction}

Multi-bend-achromat (MBA) lattices in fourth-generation storage rings can deliver diffraction-limited emittance, but the resulting high beam density intensifies intrabeam scattering (IBS) and shortens the Touschek lifetime \cite{Nagaoka2014}. To mitigate these effects, a growing number of diffraction-limited storage rings (DLSRs), including APS-U \cite{APS-U_PDR}, ALS-U \cite{Steier2017}, and HEPS \cite{Du2020,Du2025}, deliberately introduce transverse coupling and operate in a round-beam mode, thereby suppressing IBS-driven emittance growth and extending the beam lifetime. 
A popular approach to achieving this is to use skew quadrupoles and tune the betatron tunes near the difference resonance \cite{Du2020,Franchi2007,Xiao2015}.
This strategy contrasts with the operating philosophy of many third-generation sources, where coupling is typically minimized to preserve ultra-low vertical emittance \cite{Nghiem2002,Wang2015}.

The introduction of transverse coupling also brings additional considerations. When combined with a damping partition number $J_x > 1$, it can increase the total transverse emittance \cite{kuske2023}, and it may also affect nonlinear dynamics, dynamic aperture, and injection efficiency in certain lattice designs \cite{Du2020,Huang_2018}. Beyond these effects, transverse coupling can reshape collective instability through a purely linear mechanism: near the difference resonance, impedances and chromaticities are shared between eigenmodes. Lindberg analyzed this mechanism and showed that the effective wake and chromaticity governing instability become linear combinations of the uncoupled horizontal and vertical values \cite{lindbergIPAC2021,Lindberg2025}. This framework was also verified through particle tracking simulations for short-range wakefields. Whether this sharing mechanism extends to transverse coupled-bunch instabilities (CBIs) driven by long-range wakefields, including resistive-wall wakes and high-order modes (HOMs), has not yet been systematically examined. The present work focuses on the transverse HOM-driven case. HOMs are electromagnetic resonances excited by the circulating beam in storage-ring vacuum components, particularly radio-frequency (RF) cavities \cite{Marhauser_2017,Shalva2022} and in-vacuum undulators (IVUs) \cite{Li_2023, Wang2025}. When their resonant frequencies are close to beam spectral lines, HOMs can resonantly couple the motion of multiple bunches, leading to exponential growth of beam oscillations and even beam loss \cite{Ng2006}.

Transverse CBIs can be analyzed using different theoretical descriptions, depending on whether the instability growth rate is small compared with the synchrotron frequency $\omega_s$ (the weak regime) or comparable to it (the strong regime). In the weak regime, the Sacherer formalism provides a reliable description by assuming a single dominant synchrotron mode \cite{Sacherer1974}. In the strong regime, multiple synchrotron modes couple, and the more general Lindberg theory, derived from the Vlasov-Fokker-Planck equation, is required \cite{Lindberg2021}. Chromaticity plays an important role in both regimes by introducing a head-tail phase shift that can suppress instability growth through Landau damping \cite{Sacherer1974,Lindberg2021,Cullinan2016}. HOM-driven CBIs can also be studied using multiparticle tracking codes. In such simulations, HOMs can be treated as general long-range wakefields \cite{Bassi2016}, where the transverse kick on a bunch is obtained by convolving the dipole moments of preceding bunches with the wake function over multiple turns. This approach, however, requires storing the history of all prior bunch passages. A more efficient alternative for narrow-band HOMs is the voltage phasor method, which tracks complex voltage phasors on a bunch-by-bunch basis and eliminates the need for extended history storage \cite{Ruprecht2016}. We previously developed two bin-free implementations of the voltage phasor algorithm \cite{LI2026}  within the GPU-accelerated framework of the STABLE code ~\cite{He2021}, which was originally designed for longitudinal collective instabilities. With these implementations, we systematically investigated the combined effects of chromaticity, bunch length, and impedance on HOM-driven CBIs \cite{LI2026}. Those previous studies, however, were restricted to a single transverse plane without betatron coupling.

In this paper, we systematically study how the coupling-induced sharing mechanism modifies transverse HOM-driven CBI, using the HALF ~\cite{Bai2021,bai2026} storage ring as an example. The remainder of this paper is organized as follows. Section~\ref{sec:theory} presents the coupling theory, the Sacherer and Lindberg CBI formalisms, and the tracking simulation methodology. Section~\ref{sec:Numerical examples} presents simulation case studies. Conclusions are given in Section~\ref{sec:Summary}.

\section{Theory and simulation method}
\label{sec:theory}

\subsection{Theory}

\subsubsection{Betatron coupling}
\label{subsec:betatron_coupling}

Equations~(\ref{eq:coupling_coefficient})--(\ref{eq:theta}) below follow Chao's textbook~\cite{wuchao2022,Chao2020Lectures} under the weak-coupling near-difference-resonance assumption.

Linear betatron coupling is introduced by skew quadrupole magnets. When multiple skew quadrupoles are distributed around the ring, their total coupling effect can be characterized by a linear coupling coefficient:
\begin{equation}
g = \left| \frac{1}{4\pi} \int_{s}^{s+C} k^s_1(z) \, \sqrt{\beta_x(z)\beta_y(z)} \, \exp\!\left[i\left(\psi_x(z) - \psi_y(z) - (\nu_x - \nu_y - n)\frac{2\pi z}{C}\right)\right] dz \right|,
\label{eq:coupling_coefficient}
\end{equation}
where \( k^s_1\) is the local normalized skew quadrupole strength, \(\beta_{x,y}\) the betatron functions, \(\psi_{x,y}\) the betatron phase advances, \(\nu_{x,y}\) the  uncoupled tunes, \(n\) an integer for the difference resonance, \(C\) the ring circumference, and \(z\) the longitudinal coordinate. 

In the presence of coupling, the horizontal and vertical motions are no longer independent, and the transverse motion is described by two eigenmodes, denoted by \((+)\) and \((-)\). Their fractional tunes depend on \(\Delta q\) and \(g\) as:
\begin{equation}
q_{\pm}=\frac{q_x+q_y}{2}\pm\frac{1}{2}\sqrt{(\Delta q)^2+(2g)^2},
\label{eq:eigen_tunes}
\end{equation}
where \(q_{x,y}\) are the fractional parts of \(\nu_{x,y}\), and \(\Delta q = q_x - q_y\).

The degree of emittance sharing in a coupled lattice is commonly characterized by the emittance coupling factor \(\kappa_\epsilon = \epsilon_y / \epsilon_x\). The equilibrium horizontal and vertical emittances without IBS can be written in terms of \(\kappa_\epsilon\):
\begin{equation}
\epsilon_x = \frac{\epsilon_{x0}}{1 + \kappa_\epsilon / J_x}, \qquad
\epsilon_y = \frac{\kappa_\epsilon \epsilon_{x0}}{1 + \kappa_\epsilon / J_x},
\label{eq:eps_kappa}
\end{equation}
where \(\epsilon_{x0}\) is the natural horizontal emittance of the uncoupled lattice, and \(J_x = \tau_y / \tau_x\) is the horizontal damping partition number, with \(\tau_x\) and \(\tau_y\) the horizontal and vertical radiation damping times, respectively. The relationship between \(\kappa_\epsilon\), the coupling coefficient \(g\), and the fractional tune difference \(\Delta q\) is given by:
\begin{equation}
\kappa_\epsilon = \frac{g^2}{g^2 + (\Delta q)^2 / (1 + J_x)}.
\label{eq:kappa}
\end{equation}

The coupling angle $\theta$  is introduced via:
\begin{equation}
\tan 2\theta = \frac{2g}{\Delta q}, \quad 0 \leq \theta \leq \pi/2,
\label{eq:theta}
\end{equation}
with the branch choice $\theta \leq \pi/4$ for $\Delta q \geq 0$, and $\theta > \pi/4$ for $\Delta q < 0$.

Using Eq.~(\ref{eq:theta}) together with Eq.~(\ref{eq:kappa}), and eliminating $g$ and $\Delta q$, yields
\begin{equation}
\sin^2 2\theta = \frac{4\kappa_\epsilon}{(1 + 3\kappa_\epsilon) + J_x (1 - \kappa_\epsilon)}.
\label{eq:theta1}
\end{equation}
Substituting Eq.~\eqref{eq:theta1} into Eq.~\eqref{eq:eps_kappa} gives the explicit forms of the equilibrium emittances in terms of the coupling angle ~\cite{lindberg2014AOP}:
\begin{equation}
\epsilon_0 = \epsilon_{x0} \frac{1 + \frac{1}{4\tau_x}(\tau_y - 3\tau_x)\sin^2 2\theta}{1 + \frac{1}{4\tau_x\tau_y}(\tau_x - \tau_y)^2\sin^2 2\theta},
\label{eq:epsx}
\end{equation}
\begin{equation}
\epsilon_y = \epsilon_{x0} \frac{\frac{1}{4\tau_x}(\tau_y + \tau_x)\sin^2 2\theta}{1 + \frac{1}{4\tau_x\tau_y}(\tau_x - \tau_y)^2\sin^2 2\theta}.
\label{eq:epsy}
\end{equation}

A crucial consequence of the eigenmode transformation is that plane-specific parameters, including transverse short-range wakefields \(W_{x,y}\) and chromaticities \(\xi_{x,y}\), are redistributed between the two eigenmodes according to the sharing relations~\cite{lindbergIPAC2021}:
\begin{equation}
\begin{pmatrix} a_+ \\ a_- \end{pmatrix}
=
\begin{pmatrix}
\cos^2\theta & \sin^2\theta \\
\sin^2\theta & \cos^2\theta
\end{pmatrix}
\begin{pmatrix} a_x \\ a_y \end{pmatrix},
\label{eq:sharing}
\end{equation}
where \(a\) denotes any of the aforementioned parameters.

\subsubsection{Coupled-bunch instabilities}
\label{subsec:hom_instabilities}

A transverse HOM is represented by a resonator impedance of the form
\begin{equation}
Z_T(\omega) = \frac{\omega_r}{\omega}\frac{R_T}{1 + iQ\left(\frac{\omega_r}{\omega} - \frac{\omega}{\omega_r}\right)},
\label{eq:hom_impedance}
\end{equation}
where $\omega_r$ is the resonant angular frequency, $Q$ the quality factor, and $R_T$ the transverse shunt impedance with the  subscript \(T\) denoting either \(x\) or \(y\) plane. 

For uncoupled transverse motion, in the weak-instability regime ($G_\mu \ll \omega_s$), the classical Sacherer formalism is applied. For a Gaussian bunch with small chromaticity \(\xi_T\), the growth rate of the \(\mu\)-th coupled-bunch mode, which is the imaginary part of the complex angular frequency \(\Omega_\mu\) of the coherent transverse oscillation, is approximated by~\cite{Sacherer1974,Su2025}
\begin{equation}
G_\mu =\operatorname{Im}(\Omega_\mu)\approx -\frac{e I_0 f_0 \beta_T}{2E_0} \sum_{p=-\infty}^{\infty} \operatorname{Re} Z_T(\omega) \exp\!\left[-(\omega-\omega_\xi)^2\sigma_t^2\right].
\label{eq:sacherer}
\end{equation}
Here \(I_0\), \(f_0\), \(\beta_T\), \(E_0\) and \(\sigma_t\) denote the total beam current, revolution frequency, betatron function, beam energy  and rms bunch length, respectively. \(\omega_\xi = \xi_T\omega_0/\alpha_p\) is the chromatic frequency shift,  where \(\alpha_p\) is the momentum compaction factor,  and \(\omega = (pN_b + \mu)\omega_0 + \omega_\beta\), where \(\omega_\beta\) is the angular betatron oscillation frequency, \(p\) is an integer, and \(N_b\) is the total number of bunches.  

When \(G_\mu\) becomes comparable to \(\omega_s\), the instability enters the strong regime and a more general theory was developed by Lindberg \cite{Lindberg2021}.  For a Gaussian bunch, the complex angular frequency \(\Omega\) is determined by solving 
\begin{equation}
1 = -\frac{i\hat{\lambda}}{\omega_\xi\sigma_t}\sqrt{\frac{\pi}{2}}\frac{e^{-\hat{\Omega}^2/2\omega_\xi^2\sigma_t^2}}{1 - e^{2\pi i\hat{\Omega}}} \left[\operatorname{erfc}\!\left(\frac{-i\hat{\Omega}}{\sqrt{2}\omega_\xi\sigma_t}\right) + e^{2\pi i\hat{\Omega}}\operatorname{erfc}\!\left(\frac{i\hat{\Omega}}{\sqrt{2}\omega_\xi\sigma_t}\right)\right],
\label{eq:lindberg}
\end{equation}
with \(\hat{\lambda} = \frac{\lambda}{\alpha_p \sigma_\delta / \sigma_t}\) and  \(\hat{\Omega}= \frac{\Omega}{\alpha_p \sigma_\delta / \sigma_t}\) are the dimensionless eigenvalue and complex frequency, and \(\operatorname{erfc}\) is the complementary error function. Here \(\sigma_\delta\) is the rms energy spread.

A detailed comparison of the two formalisms can be found in Ref.~\cite{LI2026}, where it is also shown that for the HALF parameters, the impedance $\beta_yR_y$ levels on the order of tens of \(\text{M}\Omega\) correspond to the weak regime, while hundreds of \(\text{M}\Omega\) correspond to the strong regime.

When betatron coupling is introduced, the eigenmodes contain contributions from both transverse planes, and their effective impedances and chromaticities are determined by the sharing relations in Eq.~\eqref{eq:sharing}. These sharing relations were originally developed for collective effects driven by short-range wakefields~\cite{lindbergIPAC2021}. Here, we examine whether the same effective-parameter description can be applied to long-range HOM-driven CBIs. The coupled-bunch growth rates are calculated using the single-plane Sacherer and Lindberg formalisms with the plane-specific impedance $Z_{x,y}$ and chromaticity $\xi_{x,y}$ replaced by the corresponding eigenmode parameters $Z_\pm$ and $\xi_\pm$. The applicability of this description is assessed through comparison with macroparticle tracking in the weak- and strong-instability regimes in Sec.~\ref{sec:Numerical examples}.

\subsection{Macro-particle tracking}
The STABLE code was originally developed for simulating longitudinal beam dynamics in electron storage rings, with GPU acceleration and support for arbitrary fill patterns, passive harmonic cavities, and longitudinal HOMs \cite{He2021}. In our previous work \cite{LI2026}, we extended STABLE to include transverse dynamics by implementing two bin-free voltage-phasor algorithms for modeling transverse HOM kicks. We further extend the code with linear betatron coupling (via a thin-lens skew quadrupole model) and a transverse bunch-by-bunch feedback system (based on exponential damping), thereby enabling self-consistent macroparticle simulations of HOM-driven CBIs under betatron coupling. The implementation details are presented below.

The transverse linear motion follows a one-turn map, assuming Twiss parameters $\beta_{x0}=\beta_{y0}=1$\,m and $\gamma_{x0}=\gamma_{y0}=1$\,m$^{-1}$ at the observation point:
\begin{equation}
\begin{pmatrix}
x \\ x' \\ y \\ y'
\end{pmatrix}_{k+1}
=
\begin{pmatrix}
\cos\Psi_x & \beta_{x0}\sin\Psi_x & 0 & 0 \\
-\gamma_{x0}\sin\Psi_x & \cos\Psi_x & 0 & 0 \\
0 & 0 & \cos\Psi_y & \beta_{y0}\sin\Psi_y \\
0 & 0 & -\gamma_{y0}\sin\Psi_y & \cos\Psi_y
\end{pmatrix}
\begin{pmatrix}
x \\ x' \\ y \\ y'
\end{pmatrix}_k,
\label{eq:one_turn_map}
\end{equation}
where the phase advances are given by $\Psi_{x,y}=\Psi_{x_0,y_0}(1+\xi_{x,y}\delta)$, with $\delta = \Delta p/p$ the relative momentum deviation at turn $k$, and $\Psi_{x_0,y_0}$ the phase advance per turn at zero chromaticity (i.e., for $\delta = 0$).

For synchrotron radiation that encompasses both radiation damping and quantum excitation, the angular variable \(x'\) evolves turn by turn as
\begin{equation}
x_{k+1}' = \left( 1 - \frac{2T_0}{\tau_x} \right) x_{k}' + 2\sqrt{\frac{T_0}{\tau_x} \, \varepsilon_{x0} \, \gamma_{x0}} \cdot \text{randn},
\label{eq:synchrotron_damping_rewritten}
\end{equation}
where  $T_0$ is the revolution period, and 'randn' a random number drawn from the standard normal distribution. In the vertical plane,  the same update rule applies with \(\tau_x\) replaced by \(\tau_y\),  while the random term is omitted in the absence of quantum excitation. Consequently, the vertical emittance in the simulations arises almost entirely from betatron coupling to the horizontal plane.

We introduce betatron coupling using a single thin-lens skew-quadrupole model, following the approach in~\cite{Li2025,foosang2022} for single-bunch collective effects. With the coupling coefficient normalized as in Eq.~\eqref{eq:coupling_coefficient}, the skew-quadrupole strength \(k_1^s l\)  (with \(l\) the effective magnetic length)  is expressed as
\begin{equation}
g = \frac{1}{4\pi} (k_1^sl) \sqrt{\beta_{x0} \beta_{y0}}.
\label{eq:g_thin_lens}
\end{equation}
The corresponding transfer matrix is 
\begin{equation}
\begin{pmatrix} x \\ x' \\ y \\ y' \end{pmatrix}_{k+1} = 
\begin{pmatrix} 1 & 0 & 0 & 0 \\ 0 & 1 & -k_1^s l & 0 \\ 0 & 0 & 1 & 0 \\ -k_1^s l & 0 & 0 & 1 \end{pmatrix}
\begin{pmatrix} x \\ x' \\ y \\ y' \end{pmatrix}_{k}.
\label{eq:skew_transfer}
\end{equation}

A transverse bunch-by-bunch feedback is employed in our simulation, based on the exponential-damping model, whereas practical feedback systems are typically implemented as finite-impulse-response (FIR) filters \cite{Nakamura2024}. At turn $k$, the centroid position $\langle x \rangle_k$ and momentum $\langle x' \rangle_k$ of each bunch are computed.  The momentum of every particle in that bunch is then updated as 
\begin{equation}
x_{k+1}' = x_{k}' - \frac{2T_0}{\tau_{\mathrm{fd,x}}}
\left[
\sin\phi \ \sqrt{\frac{\beta_{\mathrm{BPM}}}{\beta_{\mathrm{kicker}}}} \langle x' \rangle_k 
- \cos\phi \ \frac{\langle x \rangle_k }{\sqrt{\beta_{\mathrm{BPM}}\beta_{\mathrm{kicker}}}}
\right],
\label{eq:fd_general}
\end{equation}
where $\tau_{\mathrm{fd,x}}$ is the horizontal feedback damping time in seconds, $\phi$ the feedback phase setting, and $\beta_{\mathrm{BPM}}$, $\beta_{\mathrm{kicker}}$ the beta functions at the BPM and kicker locations, respectively. In the following simulations, we set $\beta_{\mathrm{BPM}} = \beta_{\mathrm{kicker}}$ and $\phi = 90^{\circ}$, which corresponds to a pure resistive damper. Eq.~(\ref{eq:fd_general}) then simplifies to
\begin{equation}
x_{k+1}' = x_{k}' - \frac{2T_0}{\tau_{\mathrm{fd,x}}} \, \langle x' \rangle_k .
\label{eq:fd_direct}
\end{equation}
The same form of feedback mechanism applies to the vertical direction.

\section{Numerical examples}
\label{sec:Numerical examples}
A tentative set of parameters of HALF storage ring with insertion devices is summarized in Table~\ref{tab:table1}. We first examine linear coupling through the equilibrium emittances and eigen-tunes (Sec.~\ref{subsec:equilibrium}), then investigate how the sharing relations redistribute  chromaticity and feedback damping between the eigenmodes (Sec.~\ref{subsec:Verification}), and finally evaluate how this redistribution affects HOM-driven CBI growth rates in the weak- and strong-instability regimes (Sec.~\ref{CBI}).
\begin{table}
\begin{center}
\caption{\label{tab:table1}Main parameters of HALF}
\begin{tabular}{c c c }
\hline
Parameter & Symbol & Value \\
\hline
Ring circumference & $C$ & 479.86 m \\
Beam energy & $E_0$ & 2.2 GeV\\
Nominal beam current & $I_0$ & 350 mA \\
Harmonic number & $h$ & 800 \\
Voltage of main cavity & $V_{rf}$ & 1.2 MV \\
Energy loss per turn & $U_0$ & 380 keV \\
Momentum compaction & $\alpha_p$ & $9.4 \times 10^{-5}$ \\
Natural rms bunch length & $\sigma_{t0}$ & 7.16 ps \\
Natural energy spread & $\sigma_\delta$ & $7.44 \times 10^{-4}$\\ 
Natural horizontal emittance  & $\varepsilon_{x0}$& 60 $\rm {pm\cdot rad}$ \\
Betatron tune & $\nu_{x,y}$ & 48.15/17.15\\
Corrected chromaticity & $\xi_{x,y}$ & 5/3\\
Radiation damping time & $\tau_{x,y}$ & 17.3/20.4 ms\\ 
\hline
\end{tabular}
\end{center}
\end{table}

\subsection{Equilibrium emittances and eigenmode tunes under linear coupling}
\label{subsec:equilibrium}
To quantify the degree of transverse mixing in a coupled lattice, we adopt the emittance coupling factor \(\kappa_{\epsilon}\) as the primary independent variable. For a given \(\kappa_{\epsilon}\), the equilibrium emittances \(\epsilon_x\), \(\epsilon_y\), and their sum are given analytically by Eqs.~\eqref{eq:eps_kappa}. As evident from these equations, the equilibrium emittances and their sum depend solely on \(\kappa_{\epsilon}\) and are independent of the specific coupling strength \(g\). Accordingly, we fix \(g = 0.01\) for the subsequent emittance simulations.

In contrast to the emittances, the eigenmode fractional tunes \(q_{\pm}\) depend on both \(\kappa_{\epsilon}\) and \(g\). While \(q_{\pm}\) are formally given by Eq.~\eqref{eq:eigen_tunes} as functions of \(\Delta q\) and \(g\), the relation in Eq.~\eqref{eq:kappa} enables the elimination of \(\Delta q\), thereby expressing \(q_{\pm}\) directly in terms of \(g\) and \(\kappa_{\epsilon}\). To illustrate this dependence, we fix the vertical tune at \(\nu_{y} = 17.15\) (fractional part \(q_{y} = 0.15\)) and, for each chosen coupling strength (\(g = 0.005, 0.0075, \text{and } 0.01\)), vary \(\kappa_{\epsilon}\) by adjusting \(q_{x}\) as prescribed by Eq.~\eqref{eq:kappa}. We choose to set $q_x > q_y$, and accordingly the branch choice in Eq.~\eqref{eq:theta} gives $\theta \le \pi/4$.

In our simulations, a single bunch containing \(N_p = 10,000\) macroparticles is tracked for \(50,000\) turns using the HALF parameters listed in Table~\ref{tab:table1}, with transverse radiation damping and quantum excitation included. Coupling is implemented using a thin-lens skew quadrupole, whose strength is set according to Eq.~\eqref{eq:g_thin_lens} to obtain the target values of \(g\). The emittance coupling factor \(\kappa_{\epsilon}\) is scanned from 0.1 to 1 in steps of 0.1 for each \(g\). The equilibrium emittances \(\epsilon_x\) and \(\epsilon_y\) are calculated from the steady-state particle distribution. 
The eigenmode tunes are extracted from the FFT spectrum of the turn-by-turn vertical centroid \(\langle y \rangle\). Due to the betatron coupling, the vertical motion contains frequency components associated with both eigenmodes, which appear as two distinct peaks in the spectrum, corresponding to the fractional tunes \(q_+\) and \(q_-\).

\begin{figure}
    \centering
    \includegraphics[width=6.5cm]{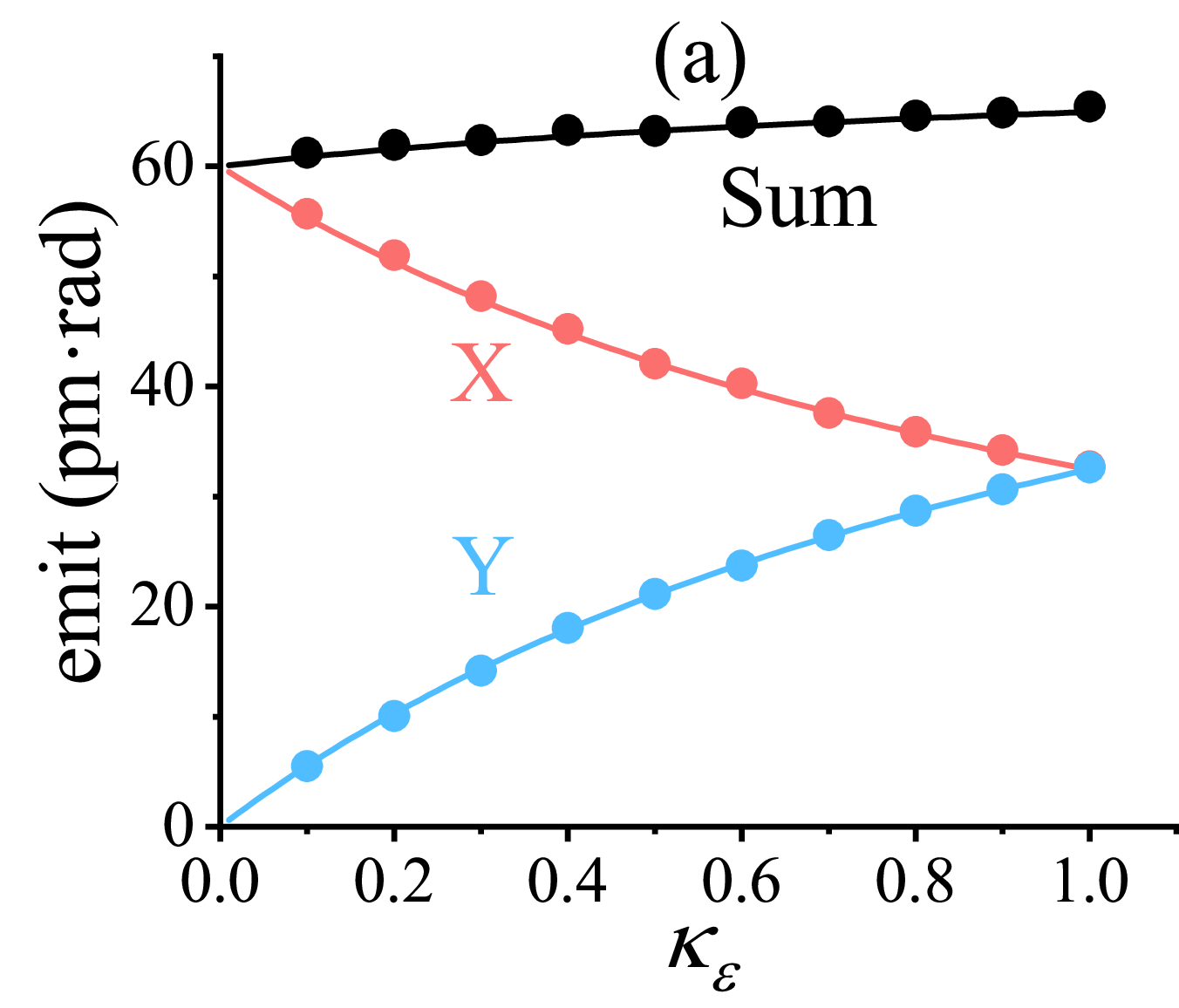}
    \includegraphics[width=6.5cm]{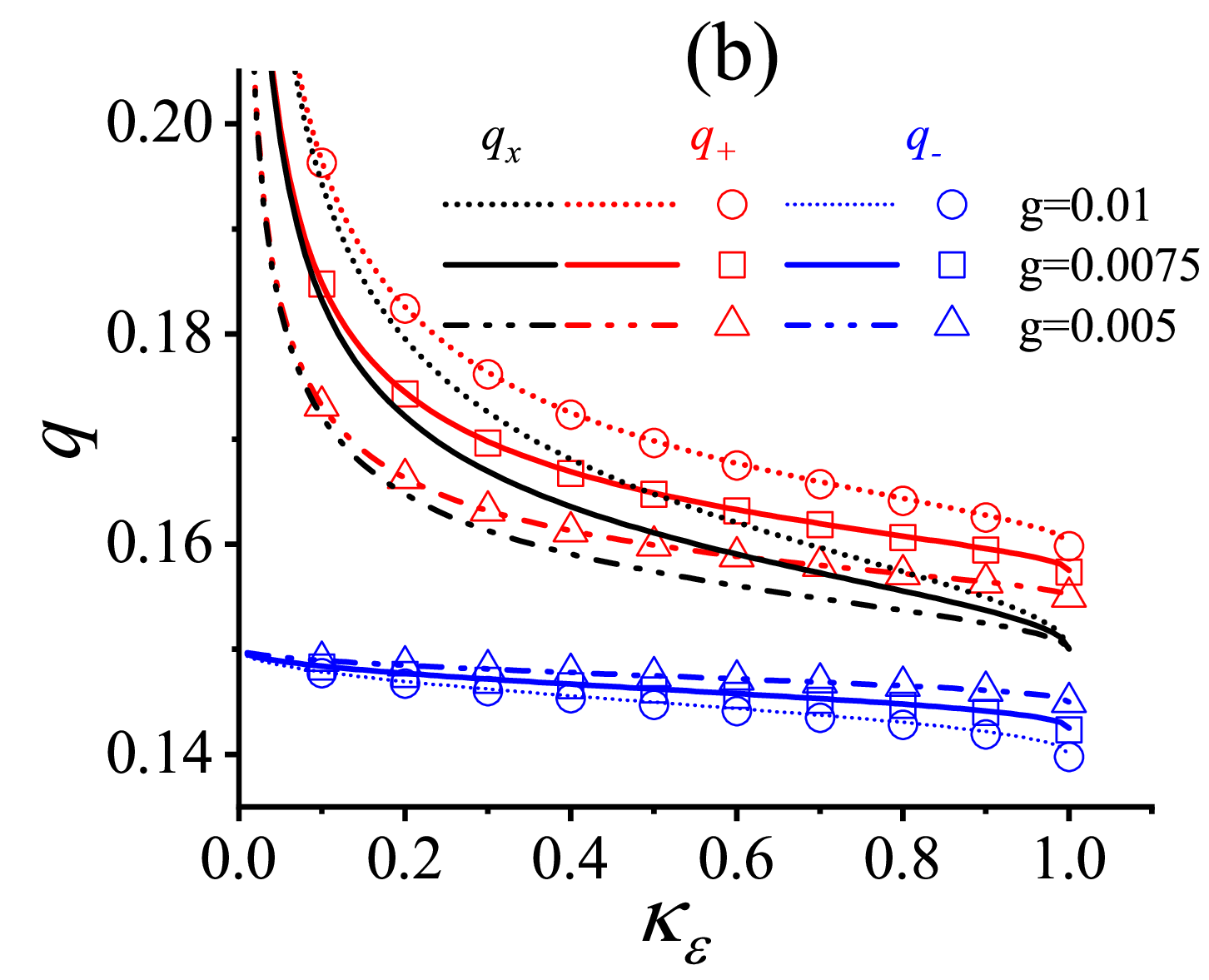}
    \caption{
(a) Equilibrium horizontal, vertical, and total emittances as functions of $\kappa_\epsilon$. Markers and lines denote simulation ($g=0.01$) and theory, respectively. 
(b) Fractional eigenmode tunes as functions of $\kappa_\epsilon$. Open markers (simulation) and lines (theory) with distinct styles correspond to different $g$. The black, red, and blue data series denote \(q_x\), \(q_+\), and \(q_-\), respectively.
}
    \label{fig1}
\end{figure}

The results are plotted in Fig.~\ref{fig1}. The excellent agreement between theory and simulation verifies the implementation of linear coupling in the tracking model. For small values of \(\kappa_{\epsilon}\), the required shift in \(q_{x}\) becomes noticeably larger for the chosen fixed value of \(g\). While a smaller \(g\) could be used to moderate this shift, the present choice is sufficient for the current demonstration.

\subsection{Sharing relations for chromaticity and feedback damping}
\label{subsec:Verification}
 
\begin{figure}
  \centering
  \includegraphics[width=6.5cm]{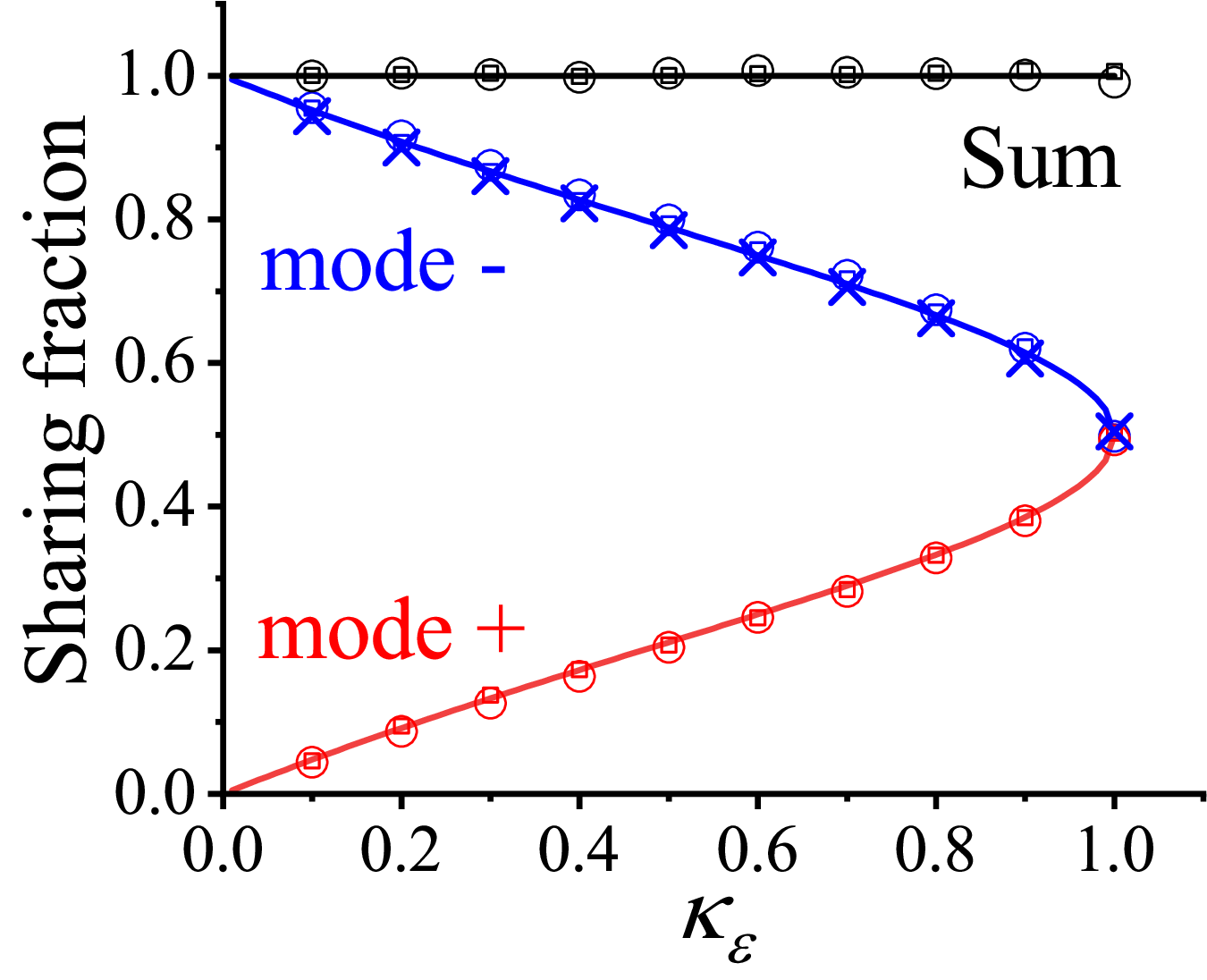}
  \caption{
Sharing fractions between the \((+)\)  and \((-)\) eigenmodes as functions of  $\kappa_\epsilon$. Theoretical predictions are shown as solid lines. Simulation results are shown as markers: chromaticity (squares), feedback damping (circles), and impedance (crosses). The black lines and markers indicate the sums of the two eigenmode fractions.
}
  \label{fig2}
\end{figure}

Before investigating HOM-driven CBIs under betatron coupling, we first examine how chromaticity and feedback damping are shared between the eigenmodes. Chromaticity affects instability suppression, while transverse feedback provides active damping, making their sharing important for interpreting the coupled dynamics. According to Eq.~(\ref{eq:theta1}), the coupling angle is determined solely by $\kappa_\epsilon$ and is independent of $g$. Therefore, we adopt $g = 0.005$ as the default coupling coefficient for all coupled simulations throughout the following sections unless otherwise stated. The analytical sharing fractions from Eq.~(\ref{eq:sharing}) are shown as solid lines in Fig.~\ref{fig2}. The impedance sharing will be discussed in the next subsection together with the CBI analysis.

To evaluate chromaticity sharing by simulation, we set $\xi_y=1$ and $\xi_x=0$. For each value of $\kappa_\epsilon$, the relative momentum deviation $\delta$ is varied from $-0.5\%$ to $+0.5\%$ in steps of $0.1\%$. The simulated eigenmode fractional tunes $q_+$ and $q_-$ are extracted at each value of $\delta$. The eigenmode chromaticities $\xi_+$ and $\xi_-$ are then obtained from linear fits of $q_+$ and $q_-$ as functions of $\delta$. The chromaticity-sharing fraction of each eigenmode is defined as $\xi_\pm/\xi_y$. The tracking results are shown as open squares in Fig.~\ref{fig2}.

We then examine the sharing of the feedback damping.  To isolate this effect, radiation damping and quantum excitation are disabled. A transverse bunch-by-bunch feedback with a damping time of $\tau_{\mathrm{fd},y}=1\,\text{ms}$ is applied solely in the vertical plane, and an initial offset of $1~\mu\text{m}$ is applied in both transverse planes.  Because the \((-)\)  eigenmode receives a larger share of the feedback damping, it decays faster, leaving the \((+)\)  mode to dominate the centroid motion. The damping rate of the  \((+)\) mode is  then extracted by fitting the exponential decay of $\langle y\rangle$, and its sharing fraction is obtained as the ratio of this fitted rate to the applied vertical feedback rate  $1/\tau_{\mathrm{fd},y}$.
To determine the rate of the \((-)\) mode, we invert the sign of $\tau_{\mathrm{fd},y}$, effectively converting damping into gain. The \((-)\)  mode then dominates, allowing its rate to be extracted from the same fitting procedure. The simulation results are shown as open circles in Fig.~\ref{fig2}.  

The simulation results for both chromaticity and feedback damping are found to be in good agreement with the analytical predictions.

\subsection{Coupled-bunch instabilities under betatron coupling}
\label{CBI}
In our previous work \cite{LI2026}, we systematically investigated transverse HOM-driven CBIs in a single transverse plane, mapping the growth rates as functions of HOM frequency, impedance strength, chromaticity, and bunch length. That study provided a baseline understanding under uncoupled conditions. Here, we extend the analysis to the betatron-coupled case. For consistency, we retain several key settings from \cite{LI2026}, as detailed in the next paragraph.

In HALF, the transverse HOM impedance is dominated by the low-frequency trapped modes of the IVUs, which are primarily in the vertical plane, with much weaker contributions from other vacuum components~\cite{He2026}. We therefore consider a representative vertical HOM with $f_r=160.85\,f_0$ (approximately $100\,\mathrm{MHz}$) and $Q=800$. Its fractional part 0.85 (equal to $1-q_y$) is resonant with a coupled-bunch mode, corresponding to the maximum growth rate. The bunch length is adjustable via a passive third-harmonic cavity. Two representative settings, $\sigma_t= 8~ \rm{ps} $ and $22.4~\rm{ps}$, are considered in this work. Each simulation case adopts one of these fixed values, and no direct comparison between different bunch lengths is made here, since the growth rate depends on the product $\xi \omega_0\sigma_t/\alpha_p$ for a Gaussian bunch, as demonstrated in our previous work \cite{LI2026}. We use a uniformly filled ring with 100 bunches in the simulations to reduce computational cost, as the growth rates are nearly identical to the 800-bunch case despite the different driven modes. To isolate the HOM-driven coherent growth, bunch-by-bunch feedback and transverse radiation effects are disabled. 
 
We first consider the weak-instability regime, with the HOM impedance set to $\beta_yR_y=20\,\mathrm{M\Omega}$ and the bunch length to  $\sigma_t=8\,\mathrm{ps}$.
For a low resonant frequency $f_r$, the most unstable mode can be estimated from Eqs.~\eqref{eq:sharing} \eqref{eq:hom_impedance}, and \eqref{eq:sacherer}    as  
\begin{equation}
\label{eq:Gmax}
G_{\max,\pm}
\approx
\frac{I_0 f_0(\beta R)_\pm}{2E_0/e}
\exp\left[
-\left(
\frac{\xi_\pm\omega_0}{\alpha_p}
\right)^2\sigma_t^2
\right].
\end{equation}
Here, $(\beta R)_\pm$ denote the effective transverse impedances for each eigenmode, obtained from the vertical-plane $\beta_yR_y$ via the sharing relation in Eq.~\eqref{eq:sharing}. As shown in Fig.~\ref{fig1}(b), for $\kappa_{\xi}\ge 0.1 $ and $g=0.005$, the deviations of $q_{\pm}$ from $q_y$  are always less than 0.024, which is much smaller than the resonance width $f_r/(f_0Q)$ ($\approx 0.2$). Thus, despite the coupling-induced frequency shift, the resonance condition remains essentially satisfied. 

As discussed above, the sharing relations in Eq.~\eqref{eq:sharing} were originally derived for short-range wakefields. Their applicability to long-range wakefields has not yet been demonstrated, although we have already used them to write down Eq.~\eqref{eq:Gmax}. We therefore first verify the impedance-sharing relation numerically. For simplicity, we set the chromaticities $\xi_{x,y}$ to zero to isolate their effects, so that the growth rate of each eigenmode is directly proportional to its effective impedance. For the vertical HOM considered here, the instability is dominated by the \((-)\) eigenmode, while the \((+)\) eigenmode signal is relatively weak and difficult to extract. We therefore focus on the \((-)\) eigenmode in the following analysis. The impedance-sharing fraction of the \((-)\) eigenmode is then obtained by
\begin{equation}
\frac{(\beta R)_-}{\beta_y R_y}
=
\frac{G_{\max,-}(\xi_{x,y}=0,\kappa_\epsilon)}
     {G_{\max,-}(\xi_{x,y}=0,\kappa_\epsilon=0)}
\end{equation}
The simulated results are added to Fig.~\ref{fig2}, where the crosses denote the impedance-sharing fraction of the \((-)\) eigenmode. For the uncoupled case $\kappa_\epsilon=0$, we set $g=0$ to recover the single-plane limit, instead of the default value $g=0.005$ used in the coupled cases. The tracking results agree closely with the theoretical sharing fraction predicted by Eq.~\eqref{eq:sharing}, confirming that the sharing relations indeed hold for the shunt impedance as well. 

\begin{figure}
  \centering
  \includegraphics[width=6.5cm]{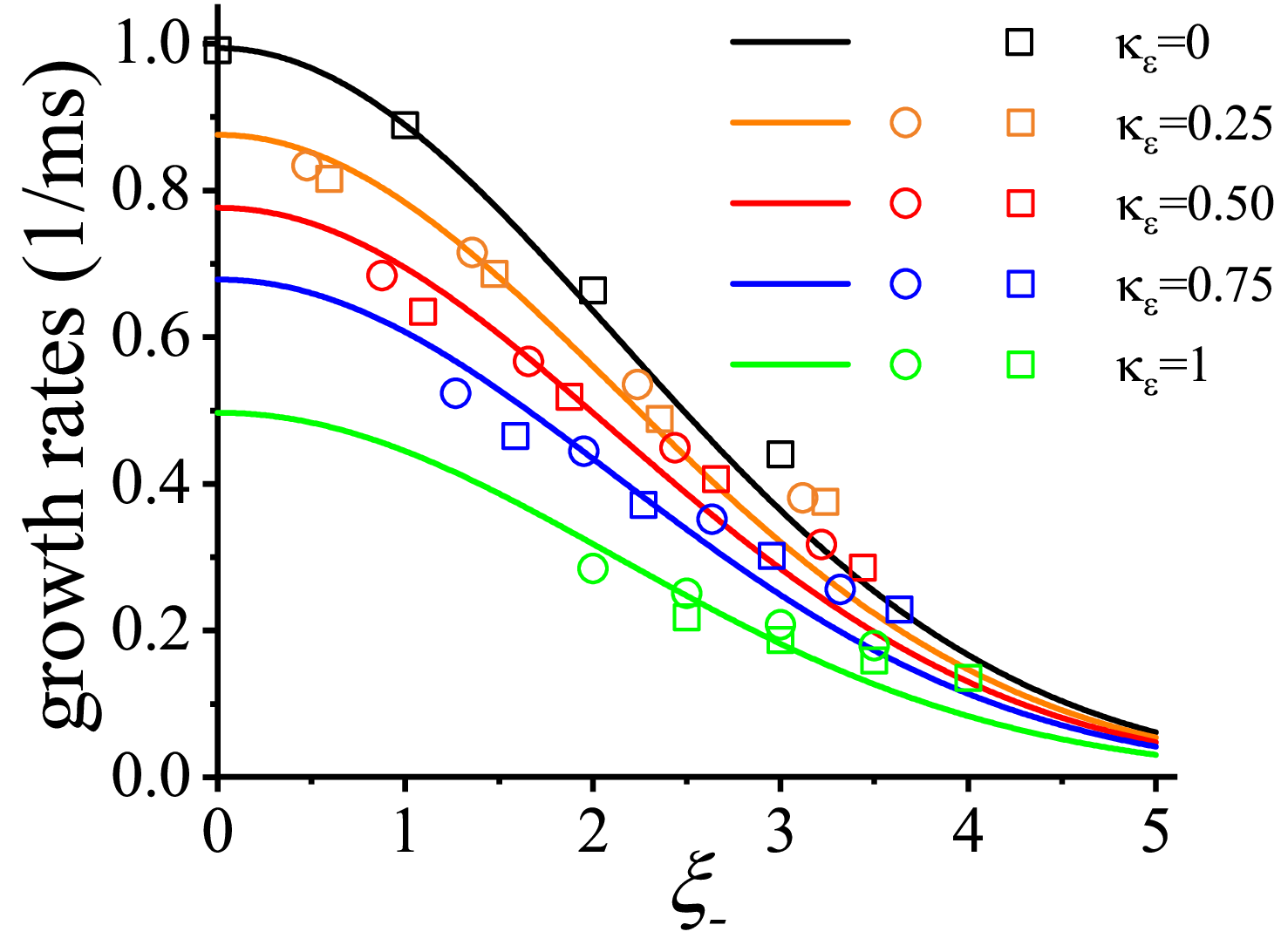}
  \caption{
Growth rates of the dominant \((-)\) eigenmode as functions of \(\xi_{-}\) for different \(\kappa_\epsilon\), with
\(\beta_y R_y=20\,\mathrm{M\Omega}\) and \(\sigma_t=8\,\mathrm{ps}\). Solid lines denote the theoretical predictions from Eq.~\eqref{eq:Gmax}. Open circles and squares denote the tracking results for \(\xi_x=4\) and \(5\), respectively, with \(\xi_y\) varied from 0 to 3 in steps of 1.}
  \label{fig3}
\end{figure}

Having established the impedance-sharing relation at zero chromaticity, we next turn to the combined effects of chromaticity and impedance under betatron coupling. Figure~\ref{fig3} shows the growth rates of the dominant \((-)\) eigenmode as functions of the chromaticity \(\xi_-\) for different values of \(\kappa_\epsilon\).  Good agreement is observed between theory and simulation for small \(\xi_-\), while for larger \(\xi_-\) values, higher-order head-tail modes must be taken into account, which is also the case in the uncoupled limit \cite{LI2026}. 

\begin{figure}
  \centering
  \includegraphics[width=6.5cm]{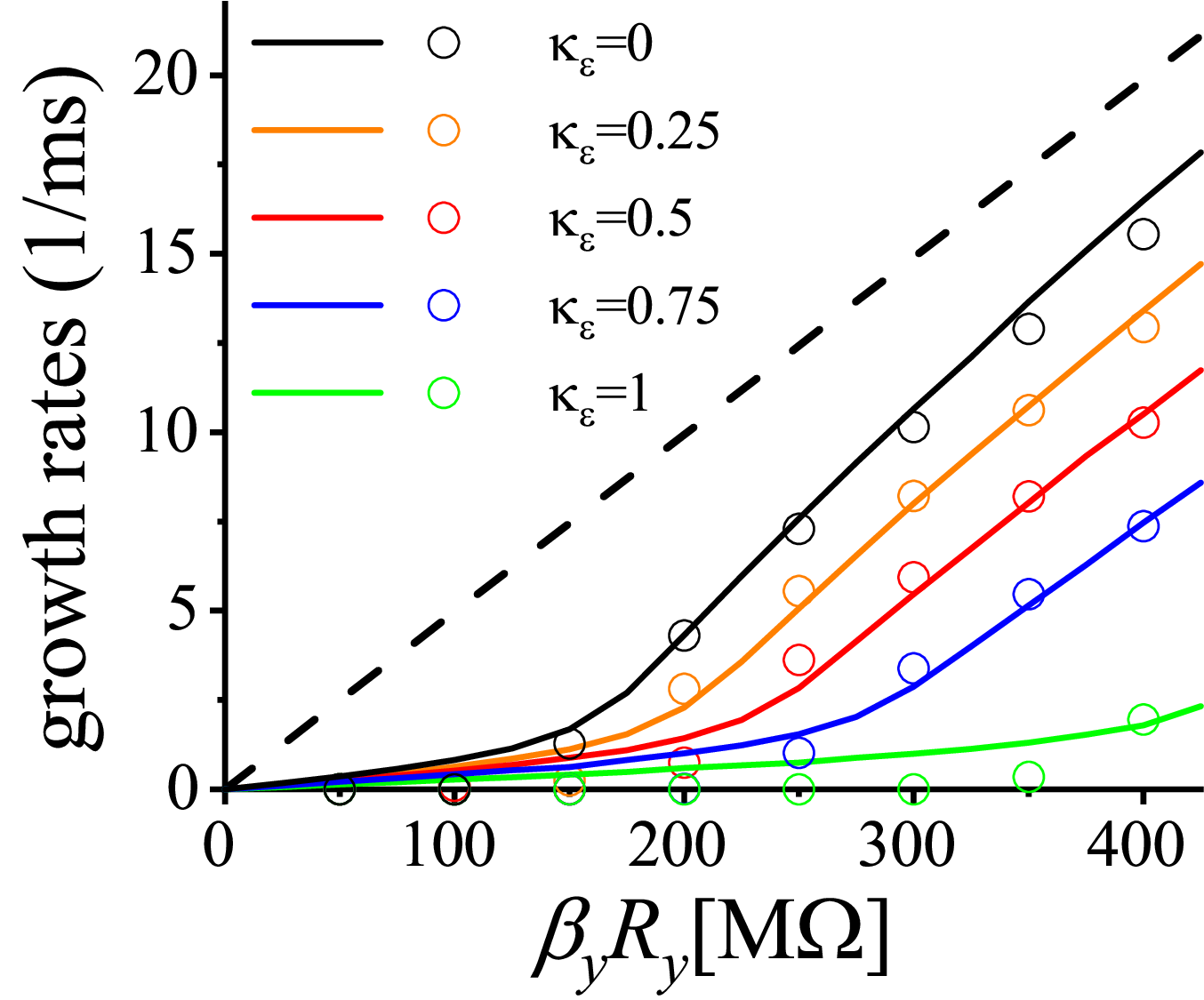}
  \caption{
Growth rate of the dominant \((-)\) eigenmode as a function of $\beta_yR_y$ in the strong-instability regime for different values of $\kappa_\epsilon$, with $\sigma_t=22.4\,\mathrm{ps}$, $\xi_x=5$, and $\xi_y=3$. Lines denote the theoretical predictions and markers denote the coupled tracking results. The dash line shows the uncoupled zero-chromaticity reference prediction with $\xi_x=\xi_y=0$ and $\kappa_\epsilon=0$.
}
  \label{fig4}
\end{figure}

We next consider the strong-instability regime, with $\beta_yR_y$ increased to several hundred $\mathrm{M\Omega}$. We set the chromaticities to $\xi_x=5$ and $\xi_y=3$, and the bunch length to $\sigma_t=22.4\,\mathrm{ps}$. In this regime, the theoretical growth rates are calculated using Eq.~\eqref{eq:lindberg}, with the impedance and chromaticity of the $(-)$ eigenmode replaced by $Z_-$ and $\xi_-$ according to the sharing relations in Eq.~\eqref{eq:sharing}. Figure~\ref{fig4} shows the growth rates as functions of $\beta_y R_y$ for different values of $\kappa_\epsilon$. The theoretical predictions agree well with the tracking results over the examined impedance and coupling ranges.

Taken together, the weak- and strong-instability results confirm that the growth rates of the eigenmodes can be predicted using the corresponding single-plane CBI theories with their effective impedance and chromaticity modified via the sharing relations Eq.~\eqref{eq:sharing}. For the vertical HOM considered here, increasing betatron coupling both dilutes the effective impedance and increases the effective chromaticity of the dominant  \((-)\) eigenmode, both of which contribute to reducing the instability growth rate.

In the strong-impedance cases, increasing betatron coupling reduces the growth rate, as discussed above. However, the growth rate can still significantly exceed the radiation damping rate, making a transverse bunch-by-bunch feedback system necessary. In the HALF design, the feedback system provides a damping time of 0.1 ms in both transverse planes, so the damping time of each eigenmode remains unchanged for different values of $\kappa_\epsilon$.  For rings with unequal horizontal and vertical feedback damping times, the coupling-induced modification of the damping rates would need to be reconsidered.

\section{Summary and discussion}
\label{sec:Summary}

In this paper, we have investigated the influence of linear betatron coupling on transverse HOM-driven coupled-bunch instabilities (CBIs), using the HALF ring as a case study. Building upon the sharing mechanism originally derived for transverse short-range wakefields~\cite{lindbergIPAC2021}, we extend its applicability to long-range transverse HOM-driven CBIs. Within this framework, the growth rate of the dominant eigenmode is evaluated using the corresponding single-plane CBI theories~\cite{Sacherer1974,Lindberg2021}, with the effective impedance and chromaticity modified via the sharing relations in Eq.~\eqref{eq:sharing}. The resulting theoretical predictions agree well with macroparticle tracking results in both the weak- and strong-instability regimes over the parameter ranges considered, demonstrating the validity of this extension.

For the HALF design, the IVUs are the primary source of vertical HOM impedance \cite{He2026}, while the horizontal chromaticity is designed to be $\xi_x = 5$. When betatron coupling is introduced, the large $\xi_x$ is partially shared via Eq.~\eqref{eq:sharing}, increasing the effective chromaticity of the $(-)$ eigenmode. Together with the dilution of the vertical impedance, this leads to a significant suppression of the vertical HOM-driven instability, effectively raising the impedance threshold and relaxing the requirements on vertical chromaticity or feedback damping. 

For the horizontal plane, since no strong horizontal HOMs have been identified in the HALF design \cite{He2026}, this case is not studied here. Nevertheless, one can expect that coupling would dilute the horizontal impedance contribution to the dominant $(+)$ eigenmode, which is beneficial, but would also reduce its effective chromaticity. The net effect would depend on the specific impedance strength and the relative magnitudes of these two competing effects.

Finally, we note that the present analysis and simulations rely solely on a linear model, thereby neglecting nonlinear effects such as second-order chromaticity.

\section{CRediT authorship contribution statement}
You Sun: Investigation, Visualization, Writing – original draft.
Weiwei Li : Investigation, Methodology, Conceptualization, Software, Writing – review \& editing.
Tianlong He: Investigation, Software. 
Penghui Yang: Investigation.
Xiaoyu Liu: Investigation.
Zhenghe Bai: Supervision, Conceptualization.

\section{Declaration of competing interest}
The authors declare that they have no known competing financial interests or personal relationships that could have appeared to influence the work reported in this paper.

\section{Acknowledgements}
This work was supported by the National Natural Science Foundation of China (No.12475158 and No.12375324 ).

%\bibliographystyle{modelnum-names}
%\bibliographystyle{elsarticle-num}
%\bibliography{refs}

%\vskip3pt

%\bio{}

%\endbio

\end{document}